\documentclass[aps,twocolumn,floatfix,superscriptaddress,showpacs]{revtex4}

\usepackage{graphicx}
\usepackage{dcolumn}
\usepackage{bm}
\usepackage{textcomp}

\begin{document}

\title{Quantum resistance metrology in graphene}

\author{A.J.M. Giesbers}
\email{J.Giesbers@science.ru.nl}
\affiliation{
High Field Magnet Laboratory, Institute for Molecules and Materials,
Radboud University Nijmegen, Toernooiveld 7, 6525 ED Nijmegen,
The Netherlands
}
\author{G. Rietveld}
\affiliation{
NMi Van Swinden Laboratorium BV, Thijsseweg 11, 2629 JA Delft,
The Netherlands
}
\author{E. Houtzager}
\affiliation{
NMi Van Swinden Laboratorium BV, Thijsseweg 11, 2629 JA Delft,
The Netherlands
}
\author{U. Zeitler}
\email{U.Zeitler@science.ru.nl}
\affiliation{
High Field Magnet Laboratory, Institute for Molecules and Materials,
Radboud University Nijmegen, Toernooiveld 7, 6525 ED Nijmegen,
The Netherlands
}
\author{R. Yang}
\affiliation{
Department of Physics, University of Manchester, M13 9PL,
Manchester, UK
}
\author{K.S. Novoselov}
\affiliation{
Department of Physics, University of Manchester, M13 9PL,
Manchester, UK
}
\author{A.K. Geim}
\affiliation{
Department of Physics, University of Manchester, M13 9PL,
Manchester, UK
}
\author{J.C. Maan}
\affiliation{
High Field Magnet Laboratory, Institute for Molecules and Materials,
Radboud University Nijmegen, Toernooiveld 7, 6525 ED Nijmegen,
The Netherlands
}

\date{\today}

\begin{abstract}
We have performed a metrological characterization of the quantum
Hall resistance in a 1~$\mu$m wide graphene Hall-bar. The longitudinal 
resistivity in the center of the $\nu=\pm 2$ quantum Hall plateaus 
vanishes within the measurement noise of 20~m$\Omega$ upto 2~$\mu$A. 
Our results show that the quantization of these plateaus is within the 
experimental uncertainty (15~ppm for 1.5~$\mu$A current) equal to that 
in conventional semiconductors. The principal limitation of the present 
experiments are the relatively high contact resistances in the 
quantum Hall regime, leading to a significantly increased noise across 
the voltage contacts and a heating of the sample when a high current is 
applied.
\end{abstract}


\maketitle


The Hall resistance in two-dimensional electron systems (2DESs)
is quantized in terms of natural constants only, $R_H = h/ie^2$ with $i$
an integer number~\cite{Klitzing}. Due to its high accuracy and reproducibility
this quantized Hall resistance in conventional 2DESs is nowadays used as
a universal resistance standard~\cite{metrology}.

Recently a new type of half-integer quantum Hall
effect~\cite{NovoselovNature, ZhangNature} was found in graphene, the purely
two-dimensional form of carbon~\cite{ReviewGeim}.
Its unique electronic properties~\cite{ReviewCastroNeto}
(mimicking the behavior of charged chiral Dirac fermions~\cite{Semenoff, Haldane})
allow the observation of a quantized Hall resistance up
to room-temperature~\cite{NovoselovScience2, Giesbers},
making graphene a promising candidate for a high-temperature
quantum resistance standard. Although the quantized resistance
in graphene around the $\nu = 2$ plateau is generally believed to be equal
to $h/2e^{2}$, up to now it has not been shown to meet a metrological
standard. In this Letter we present results of the first metrological characterization
of the quantum Hall resistance in graphene. In particular, we will address
the present accuracy of quantization (15 ppm) and the experimental conditions
limiting this accuracy.

\begin{figure}[t]  
 \begin{center}
 \includegraphics[width=6cm, angle=270]{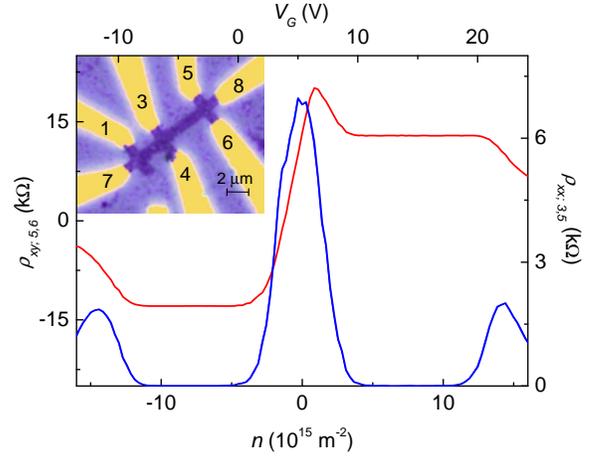}
 \end{center}
\caption{(Color online) Longitudinal resistivity  $\rho_{xx}$ (blue, measured 
across contacts 3 and 5) and Hall resistance $\rho_{xy}$ (red, measured across 
5 and 6) at $B=14$~T and $T=0.35$~K as a function of gate voltage (top x-axis) 
and the corresponding carrier concentration (bottom x-axis). A bias current 
$I=100$~nA was applied between the contacts 7 and 8. The inset shows a false 
color scanning electron micrograph of the graphene Hall-bar with the contact 
configuration of the device.
}
 \label{Figure1}
\end{figure} 

Our sample consists of a graphene Hall-bar on a Si/SiO$_{2}$ substrate forming
a charge-tunable ambipolar field-effect transistor (A-FET), where the carrier 
concentration can be tuned with a back-gate voltage $V_g$~\cite{NovoselovScience}.
In order to remove most of the surface dopants that make graphene generally
strongly hole doped and limit its mobility, we have annealed  the sample
{\sl in-situ} for several hours at 380~K prior to cooling it down slowly
($\Delta T / \Delta t < 3$~K/min) to the base temperature (0.35~K)
of a top-loading $^{3}$He-system equipped with a 15 T superconducting magnet.
After annealing, the charge neutrality point in the A-FET
was situated at 5 V and the sample displayed a (low-temperature)
mobility $\mu = 0.8$~m$^2$(Vs)$^{-1}$.

We have performed standard DC resistance
measurements using a Keithley 263 current source and two HP3458a multimeters
or, for the most sensitive longitudinal resistance measurements, an EM N11
battery-operated nanovolt meter. A low-pass {\sl LC}-filter at the current-source
output protects the sample from large voltage peaks during current reversal.
Special care was taken to achieve high leakage resistance of the wiring in
the insert ($R_{leak} > 10^{13}$~$ \Omega$).
The high precision measurements were performed with a
Cryogenic Current Comparator (CCC)~\cite{NPLCCC} using a 100~$\Omega$ transfer
resistor, where special attention was devoted to measuring at low currents
($I_{sd}=1.5$~$\mu$A).

Figure~\ref{Figure1} shows a typical quantum Hall measurement at $B=14$~T and $T=0.35$~K
with the Hall resistance $\rho_{xy}$ and the longitudinal resistivity $\rho_{xx}$
plotted as a function of the carrier concentration $n$.
Around filling factors $\nu= \pm2$ the device displays well defined flat plateaus 
in $\rho_{xy}$ accompanied by zero longitudinal resistivity minima in $\rho_{xx}$.

\begin{figure}[t]  
 \begin{center}
 \includegraphics[width=3.5cm,angle=270]{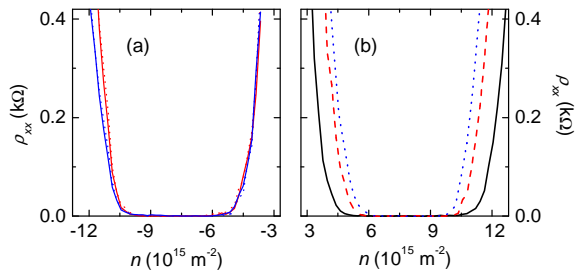}
 \end{center}
\caption{(Color online) (a) Detailed sweep of $\rho_{xx}$ for holes on both
sides of the sample, $\rho_{3,\, 5}$ (red) and $\rho_{4,\, 6}$ (blue), with
$I_{sd}=0.5$~$\mu$A at $B=14$~T and $T=0.35$~K. The curves were taken
for two different cooldowns (solid and dotted lines).
(b) Detailed sweep of $\rho_{xx;\, 4,\, 6}$ for electrons at different
source-drain currents, $I_{sd}=0.5,\, 1.5,\, 2.5$~$\mu$A, respectively
solid black, dashed red and dotted blue at $B=14$~T and $T=0.35$~K.}
 \label{Figure2}
\end{figure} 

In a next step we characterize the sample following the metrological
guidelines~\cite{Dalahaye} for DC-measurements of the quantum Hall resistance, especially
making sure that the longitudinal resistivity $\rho_{xx}$ is well enough zero
in order to provide a perfect quantization of  $\rho_{xy}$~\cite{metrology}.
Qualitatively, the absolute error in the quantization of $\rho_{xy}$ due to a finite
$\rho_{xx}$, can be estimated as $\Delta \rho_{xy} = -s \rho_{xx}$, where $s$ is in
the order of unity~\cite{Furlan}. 

In order to address the quantization conditions in some detail, we have investigated
the longitudinal resistivities in the $\nu=\pm 2$ minima along both sides of the
sample under different conditions. Figure~\ref{Figure2}(a)
shows that the $\nu=-2$ resistivity minima for holes are indeed robustly developed
on both sides of the sample for two different cooldowns. A similar
robustness of the resistivity minima is also observed for electrons around
the $\nu=2$ minimum.

Figure~\ref{Figure2}(b) displays the behaviour of $\rho_{xx}$ around
$\nu= 2$  for increasing source-drain currents. All minima remain robust
and symmetric, the position of the middle of the minimum does not change for neither
the holes nor the electrons when the bias current is increased.

\begin{figure}[t]  
 \begin{center}
 \includegraphics[width=4.5cm, angle=270]{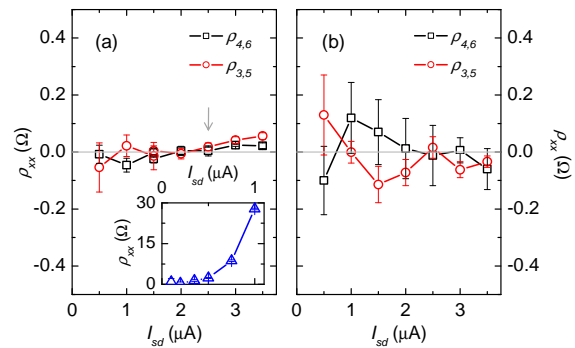}
 \end{center}
\caption{(Color online) Precise measurement of the zero longitudinal
resistance for (a) holes ($n=-7.68 \cdot 10^{15}$~m$^{-2}$), and (b)
electrons ($n=+7.89 \cdot 10^{15}$~m$^{-2}$) at $B=14$~T and $T=0.35$~K.
Current densities of 2.5~A/m for holes and 3.5~A/m for electrons are 
achievable in graphene before the quantum Hall effect starts to breakdown 
(gray arrow). The inset shows the same hole measurements for a poorly 
annealed sample. }
 \label{Figure3}
\end{figure} 

A more detailed investigation of the longitudinal resistance in its zero-minima is
shown in Figure 3.  On the hole side of the sample
(Fig.~\ref{Figure3}(a)) the resistivity in the $\nu = -2$ minimum remains
zero for bias currents  up to 2.5~$\mu$A within the measurement noise
(20~m$\Omega$ for the highest current). At higher currents the
resistance starts to rise significantly above zero, indicating current breakdown
of the quantum Hall effect.

For electrons (Fig.~\ref{Figure3}(b)), even higher currents are attainable;
no breakdown is observed for currents as high as 3.5~$\mu$A, 
corresponding to a current density of 3.5~A/m. 
For a 1~$\mu$m wide Hall bar, this is a very promising result indeed, 
as wider samples might therefore easily sustain currents 
up to several tens of microamperes
before breakdown of the quantum Hall effect starts~\cite{current}.

As a reference we also investigated a poorly annealed sample (charge neutrality
point at 9~V, mobility $\mu=0.5$~m$^2$(Vs)$^{-1}$ at 0.35~K).
Here the quantum-Hall minimum breaks down for considerably smaller currents
(see insert in (Fig.~\ref{Figure3}(a)) and already reaches 30 $\Omega$
at a current of 1~$\mu$A, making it unsuitable for high precision measurements of the QHE.

These characterization measurements presented so far are a promising starting point
to anticipate that the Hall resistance in graphene is indeed quantized accurately. From
the fact that $\rho_{xx}$ remains below 20 m$\Omega$ for currents up to 2.5 $\mu$A
one may expect an accuracy as good as 1 ppm for the quantum Hall plateaus in this well 
annealed sample.

\begin{figure}[t]  
 \begin{center}
 \includegraphics[width=4.5cm, angle=270]{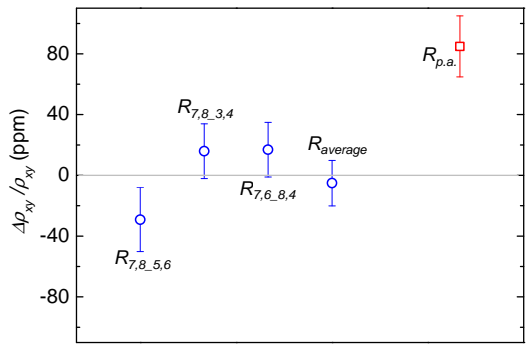}
 \end{center}
\caption{(Color online) Deviations from quantization in ppm measured with the CCC 
($I_{sd}=1.5$~$\mu$A) for different contact configurations and the average of them 
(blue circles). The red square ($R_{p.a.}$) represents the deviation for a poorly 
annealed sample at a source-drain current of 0.5~$\mu$A. }
 \label{Figure4}
\end{figure} 

In order to check this expectation we performed high precision measurements 
on the quantum Hall plateaus using a CCC with a source-drain current of 
1.5~$\mu$A (see Fig.~\ref{Figure4}). Variations measured in the quantum Hall 
resistance in a many hour CCC measurement (Fig.~\ref{Figure4}) were more 
than one order of magnitude larger than the 1 to 2 parts in $10^6$ noise 
attained in a single five minute CCC measurement run. The fluctuations in
the precision measurement are considerably reduced when better voltage contacts
are chosen. Still, the variations were two orders of magnitude larger than in a
measurement at the same current of an AlGaAs heterostructure.

Combining several measurement runs using different contacts, we achieved an
average resistance value of the $\nu=\pm 2$ quantum Hall plateaus in graphene of
$R_{H}=(12,906.34 \pm 0.20)$~$\Omega$, showing no indication of a different
quantization in graphene with respect to conventional two-dimensional electron
systems at the level of ($-5\pm15$)~parts in $10^6$.

For comparison we also determined the quantization of the poorly annealed sample
at a source-drain current of 0.5~$\mu$A. The deviation of ($85\pm 20$)~ppm is
consistent with an $s$-factor of $-0.41$ due to the finite longitudinal resistance, 
$\rho_{xx}=2.3\;\Omega$.

\begin{table}[t]
\caption{Contact resistances of the graphene sample, measured in the quantum
Hall regime where $\rho_{xx} \simeq 0$~$\Omega$ (all values for the voltage
contacts (1-6) were measured at 0.1~$\mu$A, whereas the current contacts
7 and 8 where measured at 3~$\mu$A).}
\vspace*{1em}
\centering
\begin{tabular}{c c c c}
\hline
\hline
Contact \# & $R_{holes}$~(k$\Omega$) & $R_{electrons}$~(k$\Omega$) \\
[0.5ex]
\hline
1 & 5.6  & 1.25 \\
3 & 0.95 & 6.3  \\
4 & 0.03 & 2.7  \\
5 & 1.4  & 4.8  \\
6 & 0.3  & 1.1  \\
7 & 1.0  & 5.5  \\
8 & 0.3  & 0.8  \\ [1ex]
\hline
\end{tabular}
\label{table1}
\end{table}
\vspace*{3em}

The main limitation in the CCC measurements appeared to be the contact resistance 
of the voltage contacts~\cite{Dalahaye}. The rather high resistances induce additional 
measurement noise and fluctuations in the voltage contacts thereby limiting the 
attainable accuracy of quantum-Hall precision experiments. Table~\ref{table1} shows
the contact resistances for our specific sample in the center of the $\rho_{xx}$ 
minima around $\nu= \pm 2$ in a three terminal setup. They reveal large variations 
for the different contacts, and, furthermore a significant difference between 
holes ($n<0$) and electrons ($n>0$). The latter might be explained by doping effects 
of the contacts~\cite{contacts1} and the high contact resistance of the contacts could 
be accounted for by non-ideal coupling between the gold contacts and the graphene 
sheet.~\cite{contacts2}. Besides noise on the voltage contacs, high contact
resistances also lead to local heating at the current contacts thereby limiting the
maximum breakdown current.

In conclusion, we have presented the first metrological characterization of the quantum
Hall effect in graphene. We have shown that the quantum Hall resistance in a only 1~$\mu$m wide
graphene sample is already within ($-5\pm15$)~ppm equal to that in conventional AlGaAs and
Si-MOSFET samples. A proper annealing of the sample ensuring well pronounced zeroes
in $\rho_{xx}$ and sufficiently high breakdown currents were shown to be crucial to
obtain such an accuracy. The main limitation for high accuracy measurements in our
experiments are the relatively high contact resistances of the sample used, inducing
measurement noise and local heating. 
Extrapolating our results to samples with lower resistance contacts for both
electrons and holes and using wider samples with high breakdown currents,
would most probably allow precision measurements of the quantum Hall effect in graphene
with an accuracy in the ppb range.



\end{document}